# TOWARDS 5G AIR INTERFACE TECHNOLOGY: SPARSE CODE MULTIPLE ACCESS

*Enable NOMA in Code Domain*

Fan Wei, Wen Chen, Yongpeng Wu, Jun Li, and Yuan Luo

*The fifth generation wireless networks focus on the design of low latency, high data rate, high reliability, and massive connectivity communications. Non-orthogonal multiple access (NOMA) is an essential enabling technology to accommodate the wide range of communication requirements. By coordinating the massive devices within the same resource block on power domain, frequency domain or code domain, NOMA is superior to conventional orthogonal multiple access in terms of the network connectivity, the throughputs of system and etc. Sparse code multiple access (SCMA) is a kind of multi-carrier code domain NOMA and has been studied extensively. The challenges for designing a high quality SCMA system is to seek the feasible encoding and decoding schemes to meet the desired requirements. In this article, we present some recent progresses towards the design of multi-dimensional codebooks, the practical low complexity decoder, as well as the Grant-Free multiple access for SCMA system. In particular, we show how the SCMA codebooks construction are motived by the combined design of multi-dimensional constellation and factor graphs. In addition, various low complexity SCMA decoders are also reviewed with a special focus on sphere decoding. Moreover, based on the framework of belief propagation, the SCMA Grant-Free transmission is introduced and the problem of collision resolution is also discussed.*

## Background

The Third generation partnership project (3GPP) has been evolving some new features for the next generation of wireless networks, i.e., enhanced mobile broadband (eMBB), massive machine-type communications (mMTC), and ultra-reliable low-latency communication (URLLC). The eMBB targets to provide a high data rate for the wireless networks. The typical value of fifth generation (5G) transmission rate would be 1Gbps and with the peak rate higher than 10Gbps, which is about 10 or 100 times higher than that of the fourth generation (4G) networks. In addition, to meet the massive connectivity requirement, the networks should enable millions connected devices per square kilometers in mMTC, a number far outweighs the conventional network. Moreover, URLLC necessitates the end-to-end delay should be milliseconds level in order to support the communication scenarios such as real time mobile control and vehicle-to-vehicle applications and communications.

Conventional orthogonal multiple access (OMA), such as time division multiple access (TDMA), code division multiple access (CDMA) or orthogonal frequency division multiple access (OFDMA), assigns the orthogonal resource to each user exclusively. However, with the massive connected devices, the time-frequency resources become limited and it is inefficient to allow every user to occupy the resources exclusively. Moreover, the evolving of channel codes has rendered the conventional point-to-point transmission near the Shannon limits. As a result, it is necessary to leverage the spectral efficiency, the throughputs, and the number of connected users from the perspective of network communication in the future air interface. To this end, non-orthogonal multiple access (NOMA), which allows multiple users transmit simultaneously within the shared resources becomes the uncultivated land for study.

Various NOMA techniques, such as power domain NOMA, low density signature (LDS), sparse code multiple access (SCMA), and pattern division multiple access (PDMA) had been proposed with diverse principles. While the rationales may be different, one philosophy is shared among the above techniques, i.e., there should be more than one users transmit simultaneously in each orthogonal time-frequency resource. In this article, we are going to give a comprehensive review for the recent progress on SCMA [1], a code domain NOMA. The focus would be the design of codes for SCMA with high reliability. Also, to meet the massive connectivity requirement, it is necessary to find the practical low complexity decoder for the multi-user system. To enable the low-latency communication, Grant-Free may be the alternative to conventional Request-Grant multiple access. Thus, it is interesting to consider SCMA combined with Grant-Free transmission. We would consider specifically the low complexity Grant-Free receiver design and the collision resolution problem within the contention based transmission.

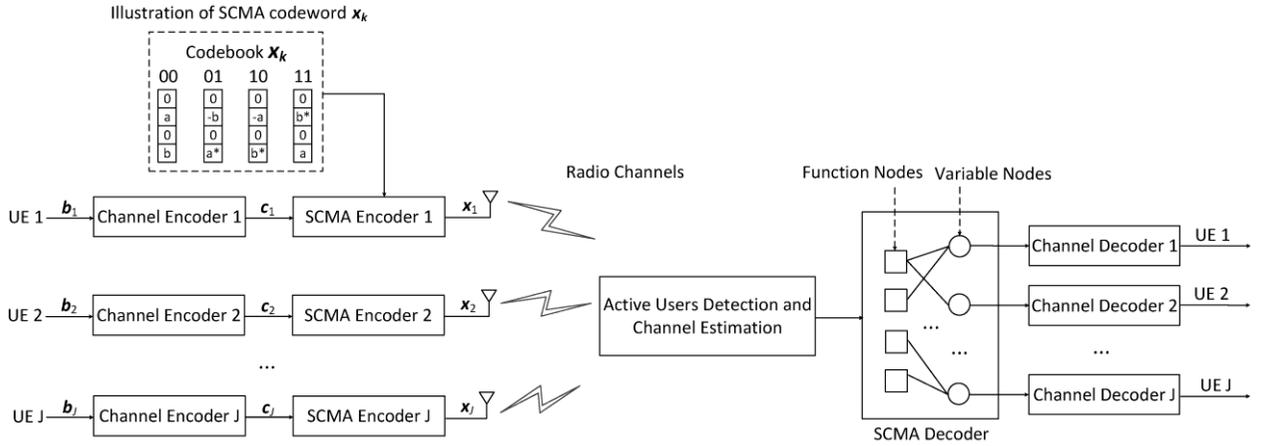

*Figure 1* The block diagram of SCMA systems.

## Sparse Code Multiple Access

In this section, we give a brief introduction for SCMA. Fig. 1 depicts the block diagram for the uplink SCMA system. In the encoder, coded bits are modulated to the multi-dimensional sparse codewords consisting of the high dimensional lattice points. As an example, the four-dimensional sparse codewords with two non-zero entries are illustrated in Fig. 2. The non-zero entries can be modulated to OFDMA subcarriers or the multiple-input-multiple-output (MIMO) antennas. Note that choosing the positions of non-zero entries can be viewed as a combinational problem, therefore at most $\binom{K}{N}$ users can be accommodated, where K is the total number of subcarriers and N denotes the number of non-zero entries. To enable the sparse structure of codewords, the number of non-zero entries should satisfy N << K. As such, a reduced number of collision users in a single subcarrier is achieved and the decoding complexity can be further decreased. For the four-dimensional codewords in Fig. 2, the maximum number of users is 6 whereas due to the sparsity of codewords, the number of collision users is reduced to 3 per subcarrier.

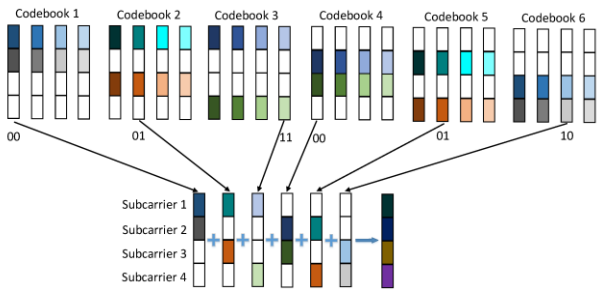

*Figure 2* Illustration of four-dimensional SCMA codebook structure.

The SCMA codebook construction can be formulated as a suboptimal multiple-stages optimization problem. A mother constellation based on high dimensional lattice points is constructed initially. Thereafter, lattice rotation for mother constellation is further applied in order to optimize the product distance of the constellation and induce the dependency and power variations among the lattice dimensions. For the subsequent codebooks design of distinct users, the constellation function operators including complex conjugate, phase rotations, and dimensional permutations for the lattice constellations are introduced. Those operators induce no alters to the minimum Euclidean distance of the constellations but only render the distinct codewords for the users.

After receiving the superimposed codewords, the decoder performs channel estimation and data decoding based on pilots as well as data signals. The active user identification is also necessary if Grant-Free transmission is considered. The sparse structure of SCMA codewords can be represented by a factor graph (FG) as in low density parity check (LDPC) codes. Motivated by this, the decoder employs message passing algorithm (MPA) for the iterative decoding of the data.

## SCMA Encoder Design and the Performance Limits Analysis

As a code domain NOMA, SCMA has the reduced average energy of signal constellation by virtue of the shaping gain of multi-dimensional codewords. Thus, the codebook design is one of the key issues of SCMA that leads to a better performance of the system. In this section, we give a review of the literatures for SCMA codebook design as well as some insights for seeking the performance limits of SCMA system.

### SCMA Codebook Design

As a multi-carrier NOMA, the codebooks of SCMA are essentially multi-dimensional constellation points. The original works of SCMA present a systematic construction method for codebook design, where the high dimensional constellation points are generated based on square-quadrature amplitude modulation (QAM). As an alternative approach, star-QAM possesses a better behavior in terms of the bit-error-rate (BER) performance in the fading channels and outperforms the square-QAM in peak-power-limited systems. Consequently, the star-QAM based codebook design for SCMA was proposed [2]. By optimizing the coordinates of the signal points in different rings of the star-QAM, the star-QAM based SCMA enjoys a larger minimum Euclidean distance compared with LDS and the codebooks in [1].

To leverage the peak to average power ratio as well as the spectrum efficiency, spherical codes were taken as another approach for the high dimensional codebook design of SCMA [3]. Various good spherical codes have been constructed through the binary codes, shells of lattices, permutations of vectors or computer searching, etc. The optimal spherical codes with low dimensions were shown to have large coding gains. By concatenating those low dimensional spherical codes, SCMA codebook with good performance can be obtained.

Apart from the mother constellation design, the factor graph of SCMA is another important factor to elevate the performance of the system. A proper designed FG may improve the system capacity, reduce the interferences of each subcarrier and result in a low decoding complexity for the receiver. In [4], the capacity of uplink SCMA is derived when the covariance matrix of the codewords is diagonal. The FG design is formulated as an integer programming optimization problem with the object function being the sum rate of the system. Each link between the variable node and function node in the FG is determined by maximizing the individual user rate in one subcarrier iteratively. In Fig. 3, the throughputs of different FG construction and power allocation schemes are compared. Through the optimization of factor graphs, it was shown that some users may occupy one subcarrier exclusively and experience no interference from other users. In other words, the irregular SCMA was shown to have a better system performance.

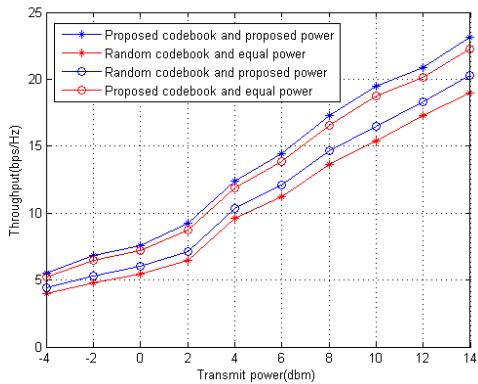

*Figure 3* Throughput comparison for different codebook and power allocation schemes.

## *Performance Limits of SCMA System*

The multi-dimensional SCMA codewords can be viewed as an equivalent MIMO system. In addition, the MPA receiver may reduce or even eliminate the multi-user interference in each subcarrier. Thus, it suffices to treat the SCMA system as a kind of single-user MIMO system. With the known results of MIMO, it is straightforward to derive the lower bound of symbol error rate for SCMA [5]. Furthermore, an inspection of the pair-wised probability of MIMO SCMA with large scale fading channel shows that the diversity order of SCMA is mainly due to the two factors, i.e., the number of received antennas and the signal-space diversity of the SCMA multi-dimensional codewords [6]. In addition to that, it was shown that the large scale fading has only the affection to the coding gain of the system.

## **SCMA Decoder Design**

The invention of turbo codes has inspired the turbo principle to the receiver design for the communication systems. Motivated by this, the iteration of soft information between the SCMA decoder and the forward error correction decoder is implemented in order to lower the coded BER of SCMA [7]. Yet with an improved BER performance, the receiver suffers high decoding complexity. Therefore, low complexity decoder is highly demanded for SCMA.

As the LDPC codes, SCMA can be represented by the factor graph. Thus, it is natural to use the belief propagation or message passing algorithm for the SCMA decoder. Due to the sparsity of the codewords, MPA has the modest decoding complexity for the low and middle modulation order systems. However, when the number of users is increasing or the modulation order is high (e.g., the 32 or 64-ary SCMA), the decoding complexity becomes prohibitively complex. As such, it is necessary to find the low complexity decoders for SCMA with satisfactory performance.

One of such decoders is based on partial marginalization [8]. The idea is motivated by the observation that the convergence rate of message passing for different users may be different. Therefore, it suffices to judge the transmit symbols for users with faster convergence rate before the algorithm termination. The judged symbols can be subtracted from the received signals and message passing will be involved with the remaining users only in the rest of the iterations. Theoretically, the complexity of partial marginalization is exponential and the performance is highly dependent on which portion of users are chosen to be judged first.

While most works focus on the low complexity decoding on the receiver side, a proper constellation design can also facilitate the low complexity computing [9]. The idea is shown in Fig. 4. For the multi-dimensional SCMA codewords with a modulation order *M*, one can project that to the constellation with $m < M$ points. Evidently, this will cause some overlaps on the point labeling. Nevertheless, the decoder is able to distinguish as long as the overlap points are separated on the other dimensions (e.g., points 01, 10 are overlapped on dimension one but are separated on dimension two). Since $m < M$, the decoding complexity is reduced from $M^{d_c}$ to $m^{d_c}$, where $d_c$ is the number of collision users in each subcarrier.

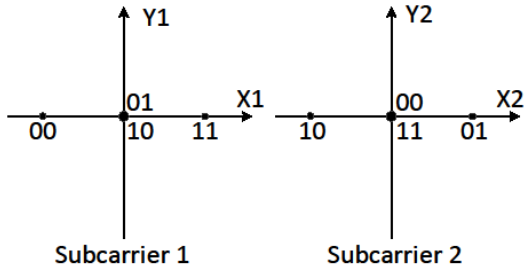

*Figure 4 Low number of projections for 4-ary SCMA.*

The dominated overheads for MPA concentrate on the function node updating. In each iteration, the MPA decoder seeks the point that is most likely to be transmitted. However, the scheme is inefficient since an exhaustive search through all possible combinations of transmitted signals has been conducted. As an alternative approach, the decoder can also proceed in a more efficient way by considering only the signal points that are not far from the received signals. The idea meets the essence of sphere decoder. For the SCMA, the decoder seeks a candidate list first by list sphere decoding (LSD). Next, MPA can be operated within the candidate list only for the reduced complexity decoding [10] [11]. The size of the candidate set can be tuned to balance the decoding complexity and the performance. Essentially, sphere decoder can be viewed as a kind of depth-first tree search algorithm. Fig. 5 shows an example of searching tree with $L=5$ levels. The search process starts from the root node and proceeds to the leaf nodes. All solid lines correspond to the survival paths during the search while the dotted lines correspond to the eliminated paths. The decoding complexity depends on the number of visited nodes on the binary tree during the search process. Thus, by avoiding the unnecessary visited nodes, a lower complexity search is possible if some branches of the tree are pruned properly. It is worth pointing out that SCMA is a kind of multi-subcarriers NOMA. As such, instead of carrying out sphere decoding on different subcarriers independently, it is beneficial to exchange the messages from different subcarriers so that some branches may be pruned in advance on the lower search levels. More specifically, when a point is reported to be excluded in the list by one subcarrier, the other subcarriers can circumvent the route lead to this point by pruning the unnecessary branches of the searching tree.

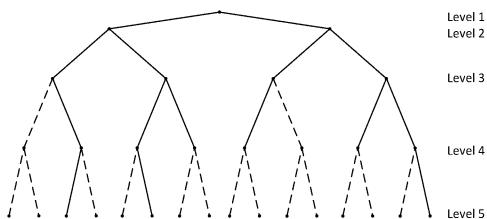

*Figure 5 Sphere decoder: depth-first tree search*

Fig. 6(a) shows the number of visited nodes with node pruning (NP-LSD-MPA) versus the conventional methods (LSD-MPA). In Fig. 6(b), it can be further observed that an improved BER performance is obtained through the node pruning. In fact, an error of LSD occurs when the actual transmit point is not contained within the candidate list. Hence, if the redundancy nodes are pruned within the list, the actual transmit point would likely to enter the final list and thus improve the performance. Sphere decoder for SCMA is also applicable for the low number of projections codewords proposed in [9], see [11] for the detailed discussion.

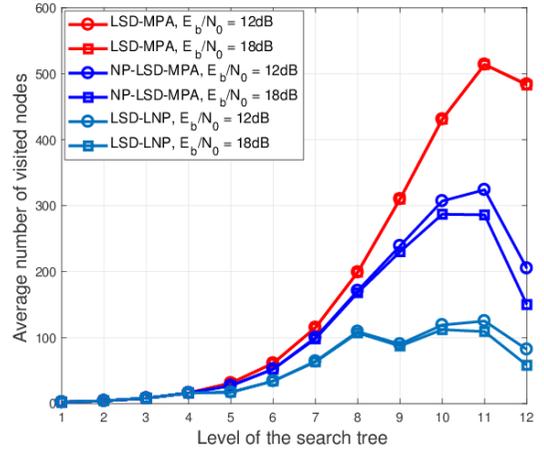

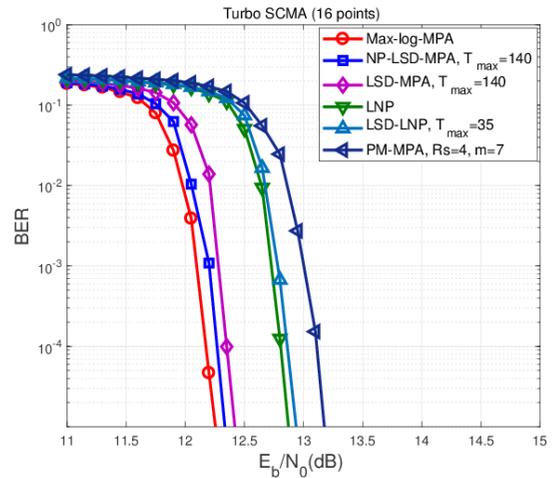

*Figure 6 An evaluation of (a) average number of nodes visited in each level of searching tree for 150% overloaded 16-ary SCMA system. LSD-MPA corresponds to list sphere decoding based decoder. NP-LSD-MPA denotes the LSD with node pruning, and LNP is the low number of projections SCMA codewords; (b) BER performance of 150% overloaded 16-ary SCMA system. The conventional max-log-MPA is used for the benchmark of performance. PM-MPA denotes the decoder based on partial marginalization [8].*

## SCMA with Grant-Free Transmission

The user scheduling issue is another problem of interest considered in the advanced multiple access techniques. In long term evolution networks, the user scheduling is realized through the Request-Grant procedure, where active users send scheduling requests to BS periodically, and wait for the grants as well as resource assignments from the base station (BS). However, the hand shakings between the BS and the active users may result in considerable delays. Meanwhile, the dynamic signaling would also degrade the spectral efficiency of the network. Thus, contention based Grant-Free transmission is advocated in the future 5G network. Instead of the hand shakings, active users contend the shared resources and transmit directly to the BS. Henceforth, two problems arise in Grant-Free multiple access, i.e., the user identification and the collision resolution.

### Active users identification for Grant-Free SCMA

For the typical 5G scenarios such as mMTC, the number of users connected to the network is supposed to be tremendous. While millions of devices are connected to the network, the simultaneously communicating users are sparse. Accordingly, it is generally to formulate the active user identification as a kind of compressed sensing problem.

Motivated by the fact the inactive users are equivalently to sending zero data symbols or having zero channel responses, the channels from entire users can be viewed as a sparse vector with nonzero entries related to the active users. Therefore, learning the support of the sparse vector is equivalent to learning the active users. In [12], by using the pilot signals solely, three algorithms are proposed to estimate the sparse channel vectors.

Generally, data-aided channel estimation would render more accurate result than the pilot-aided only one. Further, notice that user activity information are contained in not only the received pilot signals but also the data signals. Thereout, by exploiting both pilot and data signals, a receiver that performs joint user identification, channel estimation, and data decoding was proposed [13]. For the hybrid continuous (channel coefficients) and discrete (data symbols) variables estimation problem, direct maximum likelihood detection is unrealistic due to the prohibitive complexity. Thus, low complexity iterative decoder based on the factor graph of SCMA is necessary.

The factor graph representation of SCMA with hybrid variables estimation is formulated in Fig 7(a), which can be divided into three loops for the channel estimation, data detection, and user activity detection, respectively. In data detection loop, the term $p(\mathbf{x}_k|\mathbf{c}_k)$ denotes the mapping function of SCMA encoder $k$ that maps the coded bits to the SCMA codewords. The function node $f_{tn}$ corresponds to the likelihood function on the $t^{th}$ time slot, subcarrier $n$. For channel estimation loop, the tapped-delay channel model with length $L$ is considered. The function $\phi_{n,k}$ corresponds to the Fourier transformation of channel responses from the lag domain to frequency domain. To facilitate the sparse signals detection, two layer hierarchy channel model is utilized for modeling each channel tap $h_{kl}$. On the first layer, the channel is assumed to follow the Gaussian distribution with variance $\lambda_{kl}$. On the second layer, the variance is further modeled as a gamma distribution $\Gamma(\lambda_{kl}|a,b)$ with a noninformative *priori* parameter $a = 10^{-7}$ and $b = 10^{-7}$ for instance. Such a formulation of channel responses leads to the Student's t-distribution when integrating out the variables $\lambda_{kl}$. The Student's t-distribution exhibits heavy tails and thus favors sparse solutions for the active user identification problem.

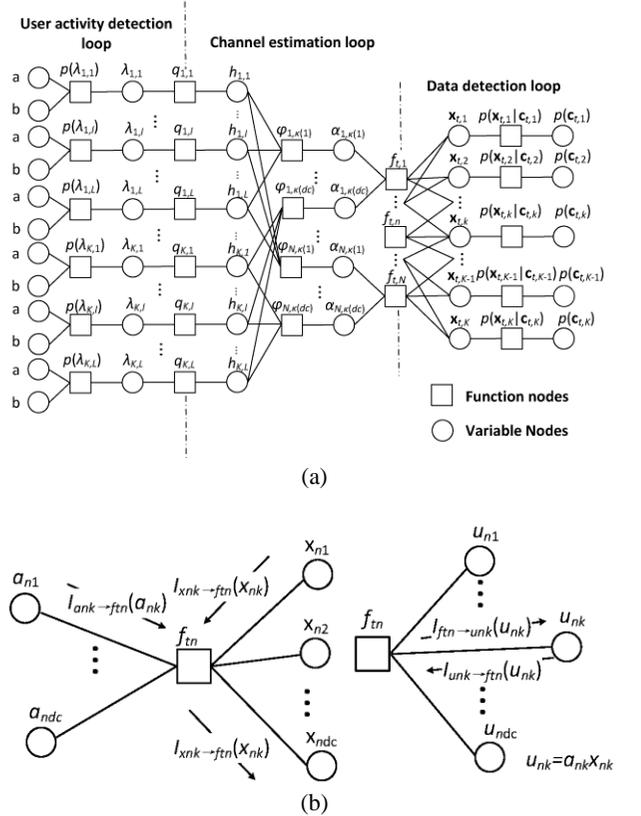

*Figure 7* Factor graph representation of (a) SCMA with hybrid variables estimation; (b) message passing within function node n. Left figure: the message passing with direct BP. Right figure: message passing with EP approximation, i.e., the extrinsic messages for virtual variable u, which is the production of channel h and data x, are approximated with Gaussian distributions with the minimized KL divergence.

With the formulation of FG, the message passing for data symbols may be calculated based on belief propagation. However, the direct BP updating is cumbersome for the hybrid continuous and discrete variables models. The idea for rendering a tractable computation is to approximate the virtual variables, defined as the production of discrete data symbols and channel coefficients, into some continuous distributions for the sake of low complexity computation. See Fig. 7(b) for the illustrations. To ensure the accuracy of approximation, the true distribution of variables is projected into Gaussian distribution for each user individually with the minimized Kullback-Leibler (KL) divergence. This coincides the idea of expectation propagation (EP) message passing. With Gaussian distributed messages, the BP updating is now tractable by some simpler computation.

Fig. 8 demonstrates the performance of receiver based on EP message passing. Machine type communications where 256 coded bits are contained in each packet is considered. The user identification comparison by the joint and pilot-aided only detector is shown in Fig. 8(a). The user activity for each user is determined by the power of the estimated channels since inactive users are equivalent to experience zero channel coefficients. It can be observed that by exploring the pilot and data symbols concurrently, the joint detector can achieve a much lower error detection rate. Fig. 8(b) compares the bit-error-rate performance for different decoding algorithms. The P-MPA algorithm that relies on pilots only for channel estimation and user detection has the worst behavior. Other joint message passing receivers, such as BP-MF and BP-GA are less effective due to their inaccurate approximation for the true probability. Owing to the probability approximation is a consequence of minimized KL divergence, the gap between the BP-GA-EP receiver [13] and the Genie that knows perfectly the channel state information and user activity information is only about 1dB.

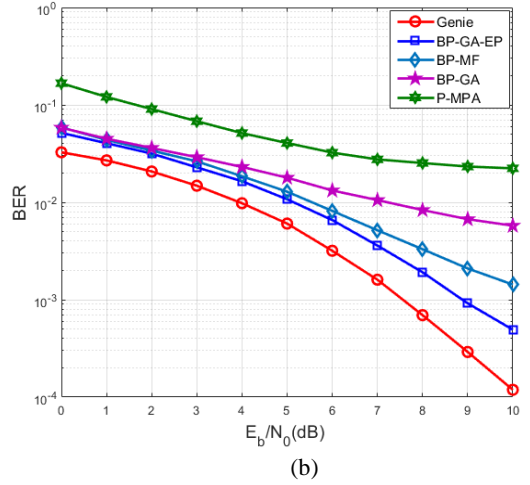

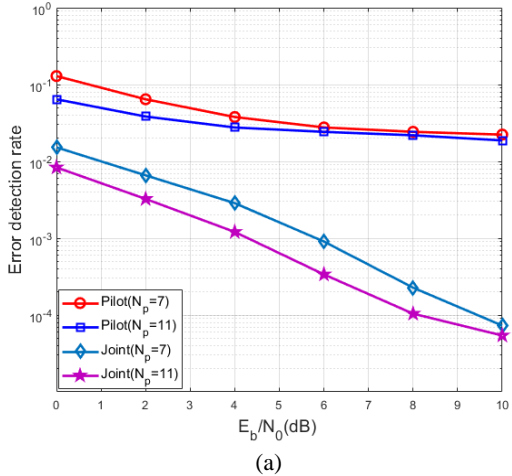

**Figure 8** *An evaluation of (a) the error detection rate by joint and pilot-only methods; (b) BER performance of 4-ary SCMA system. BP-GA-EP denotes the hybrid BP and EP message passing with Gaussian approximation. BP-MF means the hybrid BP and mean-field approximation. BP-GA is the BP Gaussian approximation using central limit theorem.*

### Collision Resolution for Grant-Free SCMA

The radio resources for Grant-Free transmission is defined in [12]. The basic radio resource, which is referred to as contention transmission unit (CTU), is defined as the combination of time, frequency, SCMA codebook and pilot. The user equipment (UE) are (possibly) mapped to CTUs via the mapping rule $CTU_{index} = UE_{ID} \mod N_{CTU}$, where $N_{CTU}$ denotes the number of CTU. In mMTC, the number of UEs would always exceed the number of available CTUs. As a consequence, multiple UEs can be mapped to the same CTU for data transmission. A collision happens when two or more active UEs transmit concurrently within the same CTU.

To resolve the UE collision problems, the conventional approach conducts a retransmission by the random back-off mechanism. However, if static mapping rule is to be utilized, constant collisions may happen when bursty traffics occur even the random back-off is employed. Further, some CTUs may be allocated with more UEs compared with the others due to the discontinuous of UE IDs. To circumvent the above issues, an ACK feedback based UE to CTU mapping rule is proposed [14]. The idea is to use the conventional static mapping rule in the first round of hand shaking between BS and UEs. Next, an ACK list containing the detection and decoding results for all the UEs is feedback from the BS to receivers. For each UE, two bits are assigned for the indicating of status: no data received (00), data successfully decoded (11) and a retransmission is needed (01), respectively. Fig. 9 is an illustration of the ACK feedback based mapping rule.

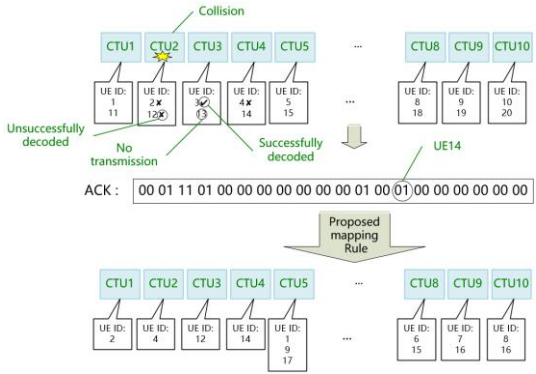

*Figure 9* Illustration of ACK feedback based mapping rule

In Fig. 10, detailed Grant-Free transmission procedure with ACK feedback based mapping rule is depicted, where transmission delay is assumed to be 3 time slots and MAP $i$ denotes the mapping rule generated in the $i^{th}$ time slot. To start with, MAP 1 in the first time slot is set based on conventional static mapping rule. Afterwards, with the receipt of ACK from BS, an adaptive MAP 4 is determined by UEs. Specifically, for UE j, if the code equals 01, i.e., the case that UE j needs to retransmit, an exclusive CTU is assigned again at the moment. The BS counts the number of CTUs associated with only one UE, i.e., $N_{single}$ and set $CTU_{index} = j \mod N_{single}$. On the contrary, if the code is not 01, an exclusive CTU is unnecessary. The BS now counts the number of CTUs associated with multiple UEs, i.e., $N_{multi}$ and set $CTU_{index} = j \mod N_{multi} + N_{single}$. The subsequent slots can be handled in a similar manner. Theoretical analysis of ACK feedback based methods shows that the collision probability would always be less than that of conventional static mapping rules.

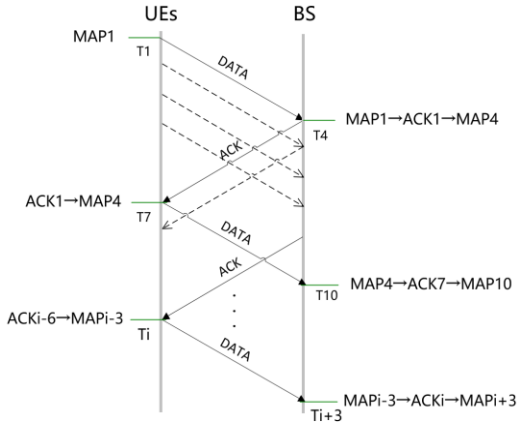

*Figure 10* Procedure of Grant-Free transmission with the ACK feedback based method

## Conclusions and Future Challenges

The state of art technologies for sparse code multiple access are reviewed in this article. We consider specifically the code domain NOMA, sparse code multiple access, as the key techniques for the future air interface. We showed how the improved SCMA codebook design can be attained through the proper mother constellation design as well as the dynamic factor graphs construction. Low complexity decoder for the SCMA with multi-user decoding is also discussed. The idea is based on of sphere decoding, where exhaustive search during the MPA is avoided. To meet the URLLC requirement, Grant-Free transmission is adapted to SCMA system. We introduced a receiver that performs joint channel estimation, data decoding, and active users detection based on EP message passing. Further, an ACK feedback based UE to CTU mapping rule is also presented in order to resolve the user collision problem.

There are several issues that require further investigations for SCMA.

- ■ The decoding of SCMA relies on the loopy BP algorithm. However, conventional BP exhibits the poor convergence behavior on the loopy factor graphs. The convergence rate is slow and algorithms could even achieve local optimal points occasionally. It is worth noting that BP rules are the consequence of minimizing the constrained Bethe free energy, which is no longer convex when FG contains loops. To this end, convex Bethe free energy needs to be constructed for loopy FG in order to overcome this problem. Further, when combined with other 5G techniques such as massive MIMO, the design of SCMA decoder should also be recast. The decoding complexity would typically grow exponentially with the number of antennas. Low complexity receiver for massive MIMO SCMA is thus of practical interest.
- ■ Since multiple users are sharing the same resource element in NOMA, the user pairing serves as another interest problem. For power domain NOMA, with fixed power allocation, the sum rate can be enlarged by selecting users with distinctive channel conditions. On the other hand, it is better to pair the users with similar channel conditions in cognitive radio inspired NOMA [15]. Unlike power domain NOMA that transmits on single resource element, the situation for SCMA is rather complex when the user paring is performed on multi-carriers.
- ■ Artificial intelligence (AI) combined with 5G is another research trend. AI can be used to solve problems that are intractable in classical communication system. For instance, combinatorial optimization is typically the NP-hard problem and is frequently encountered in SCMA such as codebook assignment, user selection, and resource allocation. Those problems can be solved alternatively by AI using the machine learning based methods. The SCMA system can also be formulated as an autoencoder, where the end-to-end communication can be joint optimized through the construction of deep neural networks. The autoencoder is trained through the BER or block-error-rate (BLER), and can accommodate various channel conditions as well as system parameters. The performance of SCMA would expected to be improved with the much lower computational complexity.

## Acknowledgment

The work is supported in part by the National Natural Science Foundation of China under Grant 61671294 and 61701301, in part by the STCSM Project under Grant 16JC1402900 and Grant 17510740700, in part by the Natural Science Foundation of Guangxi Province under Grant 2015GXNSFDA139037, and in part by the National Science and Technology Major Project under Grant 2017ZX03001002-005 and Grant 2018ZX03001009-002.